\documentclass[manuscript,screen,nonacm]{acmart}

\usepackage{url,hyperref,lineno,microtype,subcaption}
\usepackage{braket}

\usepackage{listings}
\usepackage{xcolor}
\usepackage{amsmath, xparse}

\definecolor{codegreen}{rgb}{0,0.6,0}
\definecolor{codegray}{rgb}{0.5,0.5,0.5}
\definecolor{codepurple}{rgb}{0.58,0,0.82}
\definecolor{backcolour}{rgb}{1,1,1}

\lstdefinestyle{mystyle}{
    backgroundcolor=\color{backcolour},   
    commentstyle=\color{codegreen},
    keywordstyle=\color{magenta},
    numberstyle=\tiny\color{codegray},
    stringstyle=\color{codepurple},
    basicstyle=\ttfamily\footnotesize,
    breakatwhitespace=false,         
    breaklines=true,                 
    captionpos=b,                    
    keepspaces=true,                 
    numbers=left,                    
    numbersep=5pt,                  
    showspaces=false,                
    showstringspaces=false,
    showtabs=false,                  
    tabsize=2
}

\AtBeginDocument{%
  \providecommand\BibTeX{{%
    \normalfont B\kern-0.5em{\scshape i\kern-0.25em b}\kern-0.8em\TeX}}}

\begin{document}

\title{Brain-Inspired Physics-Informed Neural Networks: Bare-Minimum Neural Architectures for PDE Solvers}

\author{Stefano Markidis}
\email{markidis@kth.se}
\affiliation{%
  \institution{KTH Royal Institute of Technology}
  \city{Stockholm}
  \country{Sweden}
}

\renewcommand{\shortauthors}{S. Markidis}

\begin{abstract}
Physics-Informed Neural Networks (PINNs) have emerged as a powerful tool for solving partial differential equations~(PDEs) in various scientific and engineering domains. However, traditional PINN architectures typically rely on large, fully connected multilayer perceptrons~(MLPs), lacking the sparsity and modularity inherent in many traditional numerical solvers. An unsolved and critical question for PINN is: What is the minimum PINN complexity regarding nodes, layers, and connections needed to provide acceptable performance? To address this question, this study investigates a novel approach by merging established PINN methodologies with brain-inspired neural network techniques. We use Brain-Inspired Modular Training~(BIMT), leveraging concepts such as locality, sparsity, and modularity inspired by the organization of the brain. With brain-inspired PINN, we demonstrate the evolution of PINN architectures from large, fully connected structures to bare-minimum, compact MLP architectures, often consisting of a few neural units!

Moreover, using brain-inspired PINN, we showcase the spectral bias phenomenon occurring on the PINN architectures: bare-minimum architectures solving problems with high-frequency components require more neural units than PINN solving low-frequency problems. Finally, we derive basic PINN building blocks through BIMT training on simple problems akin to convolutional and attention modules in deep neural networks, enabling the construction of modular PINN architectures. Our experiments show that brain-inspired PINN training leads to PINN architectures that minimize the computing and memory resources yet provide accurate results.
\end{abstract}

\keywords{Brain-Inspired PINN, Bare-Minimum PINN Architectures, Spectral Bias Phenomenon, Modular PINN}

\maketitle

\section{Introduction}
Scientific Machine Learning (SciML) is a discipline harnessing machine learning methods, such as neural networks (NNs)~\cite{goodfellow2016deep} and operators~\cite{lu2021learning}, to solve scientific computing problems, including scientific simulations, linear and non-linear solvers, inverse problems, and equation discovery~\cite{karniadakis2021physics}. One of the most active research areas in SciML is the development of Partial Differential Equations (PDE) solvers that are the backbone of scientific simulations. The SciML PDE solvers are part of the so-called SciML approaches with \textit{learning bias} as the PDE  is embedded into the loss function, and the solution is determined by the NN training or learning. This is opposed to SciML approaches with \textit{inductive bias} where the given knowledge about the modeled system, e.g., symmetries and conservation laws, influences the NN architecture design.

The core concept of SciML PDE solvers revolves around encoding the governing PDE equation into the NN loss function, facilitating numerical differentiation on NN graphs through automatic differentiation, and optimizing the loss function using techniques like the first-order Adam~\cite{kingma2014adam} or second-order BFGS~\cite{liu1989limited} optimizers. Prominent SciML PDE solvers include Physics-Informed Neural Networks (PINN)~\cite{raissi2019physics}, deep Galerkin, and Ritz methods~\cite{sirignano2018dgm,yu2018deep}. PINNs have rapidly evolved and found wide-ranging applications from computational fluid dynamics to material science and chemistry and are the focus of this work.

These PDE solvers, termed physics-informed, encode physics conservation laws into the loss function to guide the learning process toward its minimization. Conceptually, PINN serves as an extension and non-linear version of Finite Element Methods (FEM)~\cite{zienkiewicz2005finite}, with non-linear activation functions acting as piece-wise basis functions and the loss function representing the residual or equation error at each solver iteration. Its analogy and equivalence with kernel regression and other traditional methods have facilitated investigations into PINN's numerical properties, such as consistency and convergence aspects~\cite{wang2022and,mishra2023estimates}, including the PINN spectral bias (higher convergence rate for low-frequency solution components). An increased understanding of the PINN fundamental numerical properties allows us to develop numerical methods further and integrate new SciML techniques into traditional approaches~\cite{markidis2021old}. 

Despite the PINN efficacy, there is a gap in understanding the impact of PINN architecture and the development of minimal architectures in terms of neural units, layers, and connectivity while ensuring accurate results. Most PINN studies analyze fully connected multi-layer perceptron (MLP) architectures with minimal sparsity and lack of structural modularity. In contrast, traditional numerical solvers, like finite-difference linear solvers, exhibit high structure and sparsity, which is evident in scenarios such as sparse matrices arising from Poisson equation discretization. The absence of macroscale structure in PINN poses challenges for interpretability and explainability, obfuscating the manifestation of intrinsic PDE nature within the network architecture. 

In the search for PINN solvers and their building blocks that are resource-efficient and compact, we explore the application of brain-inspired neural network techniques~\cite{hassabis2017neuroscience,lillicrap2020backpropagation}. These techniques draw loose inspiration from models of brain computations, mapping neurons and synapses to neural networks and weights/biases. A key distinction between traditional and brain-inspired NN architectures lies in the learning rule, typically local instead of global back-propagation, and plasticity, reflecting the adaptivity and dynamicity of connection strengths and connectivity during training. Traditional NN architectures lack the concept of locality, whereas brain-inspired NN architectures prioritize it, leading to modularity, with specific parts of the network specialized in distinct operations. Moreover, brain-inspired NNs tend to exhibit high sparsity.

Brain-Inspired Modular Training (BIMT) is one of the most successful brain-inspired neural network techniques~\cite{liu2024seeing,liu2023growing}. Its fundamental idea introduces neural network locality by associating a spatial coordinate with each neural unit, enabling the rearrangement of neuron positions to enhance locality and modularity. This approach trains the network for increased locality by adding a loss function that penalizes non-local connections.

An important aspect concerns the number of resources that should be used to solve a given PDE with PINN. Typically, a few layers and tens of neural units, all fully connected, are used for solving simple PDE problems. Still, no insights exist on the bare minimum PINN architecture in terms of computing units capable of producing the PDE solutions.  This study addresses the fundamental research questions: \textit{What is the minimal or reduced number of PINN neural units required to produce PDE solutions using PINN?} The answer might surprise many: simple PDE requires only a few neural units to encode the PDE solution. We call this PINN \emph{bare-minimum architectures}, as they consist only of a few neural units and are only affected by a small loss of accuracy when compared to large fully-connected MLP. Through BIMT, we can derive bare-minimum architectures for solving given PDEs. We demonstrate that the spectral bias phenomenon~\cite{karniadakis2021physics} (also called F-principle~\cite{xu2019frequency}) manifests in the bare-minimum architectures: solving high-frequency problems requires more neural units than PINN solving problems with low-frequency components. Finally, we combine bare-minimum PINN architectures into modular and compact PINN architectures. These basic bare-minimum modules resemble other common deep-learning building blocks like convolutional kernels/filters and attention modules.  We show that modular PINN provides promising results in terms of performance. 

\begin{figure}
\centering
\includegraphics[width=\textwidth]{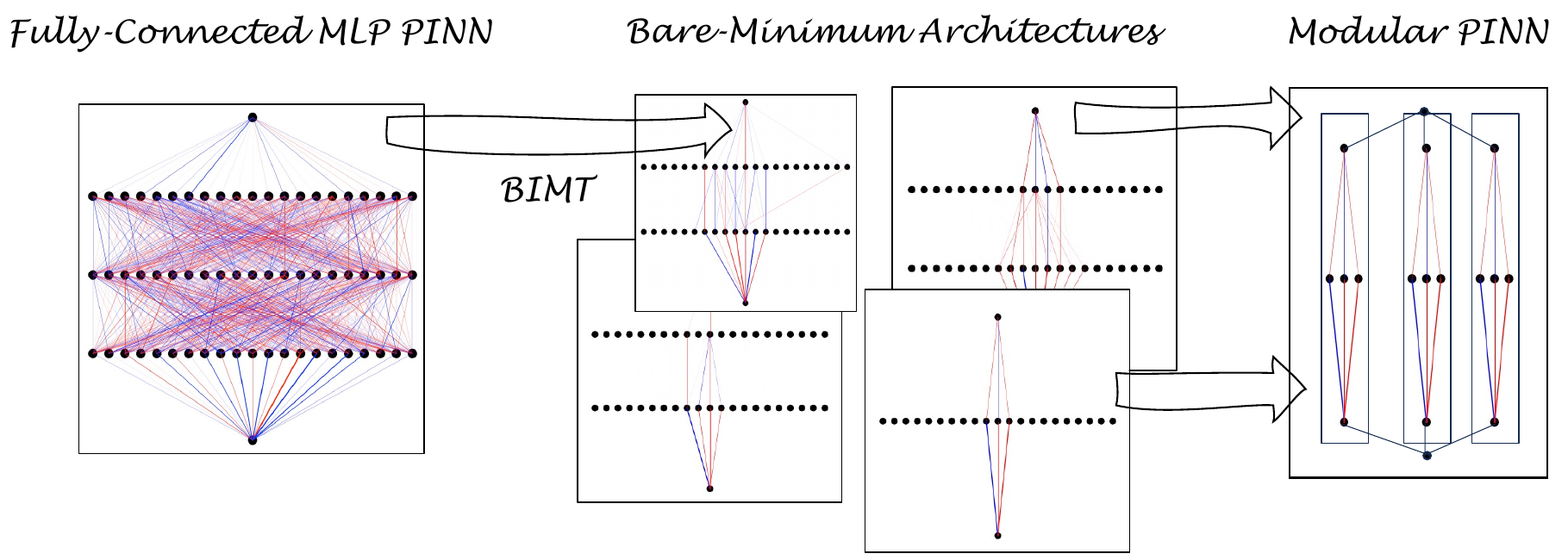}
\caption{A graphical illustration of the contributions of this work: \emph{i)} Brain-Inspired Modular Training (BIMT)~\cite{liu2024seeing}  allows us to obtain bare-minimum PINN architectures \emph{ii)}  we use bare-minimum architectures as basic modules to be combined to  build compact modular PINN architectures.} \label{fig:teaser}
\end{figure}

The primary objective of this study is illustrated in Fig.~\ref{fig:teaser}: first, we derive PINN bare-minimum architectures from fully connected MLP; second, we use bare-minimum architectures to build modular PINN. Overall, we demonstrate that BIMT leads to dynamically changing architectures with highly sparse and modular features akin to traditional numerical approaches, albeit with a slight compromise in accuracy. Furthermore, employing brain-inspired PINNs, we derive primitive building blocks that can be reused to construct larger PINNs, referred to as modular PINNs.

The contributions of this work can be summarized as follows:
\begin{itemize}
\item We demonstrate that BIMT leads to bare-minimum and compact architectures for PINN PDE solvers. For instance, we illustrate that PINN for solving simple differential equations, such as the logistic equation, requires only one neural unit in the hidden layer.
\item We identify several neural network architecture primitives capable of serving as modules in larger neural networks. By examining different basic primitives for solving the Poisson equation with source terms of increasing higher frequencies, we observe a manifestation of the PINN spectral bias on the number of PINN connections needed to represent accurately a solution, indicating that higher-frequency source terms necessitate denser PINN architectures.
\item We develop and implement a modular PINN architecture based on the identified building blocks from training brain-inspired neural networks. Our results demonstrate that modular PINN architectures built on PINN primitive modules exhibit lower test errors than fully connected MLP PINNs utilizing the same number of neural units.
\end{itemize}

The remainder of this paper is organized as follows. In Section~\ref{background}, we provide background information on PINNs and introduce the concept of brain-inspired neural networks. Section~\ref{method} details the methodology employed in this study, including the implementation of PINN based on BIMT and the experimental setup. Section~\ref{results} presents the results of our investigation, including the evolution of PINN architectures, comparisons with analytical solutions, and the derivation of modular PINN architectures. Section~\ref{related} discusses previous work in related areas of PINNs and BIMT. Finally, Section~\ref{conclusion} concludes the paper with a summary of key findings and suggestions for future research directions.

\section{Background}
\label{background}
PINNs are neural networks that take a point $t_i$ in the equation domain (referred to as the collocation point) as input and provide the approximated solution $\tilde{x}(t_i)$. The solution is encoded into the neural network with $l$ layers. At any given time, the PINN network acts as a surrogate solver, providing the approximated solution $\tilde{x}$ by running:

\begin{equation}
\tilde{x}(t_i) = a \circ Z_l \circ \ldots \circ a \circ Z_2 \circ a \circ Z_1(t_i),
\label{forward}
\end{equation}

where $\circ$ denotes the composition operation, $a$ represents the non-linear activation function, and the affine-linear maps $Z_i$ are expressed as:

\begin{equation}
Z_i (t_l)= W_i t_l + b_i.
\end{equation}

Here, $W_i$ and $b_i$ are the weights and biases of layer $i$, respectively.

The training process aims to determine the weights and biases through iterative steps. It begins with a forward pass, where Eq.~\ref{forward} is applied to several collocation points, typically chosen at random positions or according to a specific distribution. At each iteration, the forward pass yields an approximate solution at the collocation point, $\tilde{x}(t_i)$. Subsequently, the error or loss function is measured to adjust the weights and biases, a process known as \textit{back-propagation}. An optimizer then modifies the weights and biases of different layers to minimize the loss function.

The basic training process of PINNs is unsupervised, as it does not necessarily require labeled data, such as solutions obtained from other simulation techniques. This is made possible by encoding the differential equation and boundary conditions into the loss function, utilizing the residual to guide the training process. This approach is akin to Krylov subspace solvers, where the residual is minimized iteratively.

For instance, consider solving a Poisson equation $d^2 x(t) / dt^2 = \sin(t)$. The residual at a certain point $t_i$ can be calculated as:

\begin{equation}
r_i = \frac{d^2 \tilde{x}}{dt^2}\Bigg|_i - \sin(t_i).
\end{equation}

The second-order derivative at point $t_i$ is computed using automatic differentiation, which allows for calculating derivatives on the neural network, exploiting the chain rule. Unlike finite difference differentiation, automatic differentiation enables derivative calculation at any given collocation point without requiring a grid or associated spacing.

Without boundary conditions, PINNs converge to one of the infinite solutions. We impose two boundary conditions to obtain a unique solution for our second-order PDE. For example, with boundary conditions $x(0) = 0$ and $x(2 \pi) = 0$, two additional residuals are introduced at the boundary collocation points:

\begin{equation}
r_{BC_0} = \tilde{x} (0), \quad r_{BC_1} = \tilde{x} (2 \pi).
\end{equation}

These residuals are incorporated as Mean Squared Error (MSE) into the PINN loss function, which includes the following terms:

\begin{equation}
\mathcal{L}_{PINN} = \frac{1}{N} \sum_i^N  r_i^2 + \frac{1}{N{BC0}} \sum_i^{NBC0} r_{i,BC0}^2 + \frac{1}{N_{BC1}} \sum_i^{NBC1}  r_{i,BC1}^2,
\label{loss2}
\end{equation}

where $N_{BC0}$ and $N_{BC1}$ represent the number of samples taken at points $0$ and $2 \pi$, respectively. This approach serves as a soft constraint on solving a multi-learning task, minimizing the residual both within the system and at the boundary. In scenarios involving experimental or simulation data, an additional component can be added to the loss function to consider the error between the neural network and observational data. However, in this study, we adopt a fully unsupervised approach without utilizing experimental data.

\section{Brain-Inspired Physics-Informed Neural Networks}
\label{method}
This work employs the BIMT approach to determine the PINN architecture, as detailed in Ref.~\cite{liu2024seeing}. For clarity, we summarize the key features of BIMT:

\begin{itemize}
\item \textbf{L1 Penalty (Lasso Regularization)}: BIMT utilizes L1 penalty or Lasso regularization during network training to prevent overfitting and enhance generalization. This regularization induces sparsity in the weight matrix, wherein some weights may become exactly zero during training. The Lasso regularization employed in BIMT aims to increase the sparsity and modularity of the neural network architecture. A hyperparameter $\lambda$ governs the strength of the L1 penalty and can be adjusted during the simulation.
\item \textbf{Introduction of Geometry and Distance}: BIMT incorporates the notion of geometry and distance into the neural network architecture by associating a coordinate with each neural unit across different layers. In this work, we adopt a two-dimensional Euclidean space, with the $x$-direction spanning along the neural units within the same layer (input, hidden, and output layers) and the $y$-direction spanning across layers. The distance $d_{i,j}$ between two neural units is leveraged to scale the weights and biases when computing the L1 penalty. By integrating this technique, we can optimize for increased locality or minimize the distance between neural units by incorporating a loss function component that accounts for total distance. A hyperparameter $A$, related to the network size, regulates the importance of locality. For $A=0$, the L1 penalization does not consider locality.
\item \textbf{Neural Unit Swapping}: BIMT permits swapping different neural units within the same layer if it enhances locality, i.e., decreases the distance between neural units. BIMT introduces the concept of neural unit importance, computed as the sum of input and output weights to determine which neural units to swap and select critical swaps. This importance metric is utilized to swap the most significant neural units within the layer if it improves locality. However, this operation incurs computational overhead and is typically not performed at every iteration.
\end{itemize}

Consequently, BIMT yields neural network architectures that dynamically evolve during training, characterized by high sparsity due to Lasso penalization and local structure due to the additional penalty on distances between neural units. These features align with principles of brain-inspired computing, albeit resulting in slightly reduced performance (in terms of training and test accuracy) compared to fully connected MLP counterparts~\cite{liu2024seeing}.

\subsection{Implementation}

The implementation of brain-inspired PINNs extends the Python and PyTorch~\cite{paszke2019pytorch} BIMT implementation~\cite{liu2024seeing} to solve differential equations. We leverage the PyTorch \texttt{autograd}~\cite{paszke2017automatic} for automatic differentiation.

Unless otherwise specified, our implementation initiates with 21 neural units per layer, starts with a fully connected network comprising one or two hidden layers, and executes 100,000 epochs. We employ 1,000 collocation points within the domain and 50 boundary points. The weights are initialized using Xavier initialization~\cite{kumar2017weight}, and biases are set to a constant value of 0.01. The learning rate is fixed at 0.002.

Among various activation functions tested, the \texttt{sinLU} activation function~\cite{paul2022sinlu} ($a(x) = x \sin (x) \sigma(x)$, where $\sigma(x)$ is the sigmoid function) yields the most compact brain-inspired architecture. We utilize the AdamW optimizer~\cite{loshchilov2017decoupled}, which decouples weight decay and gradient update, as it provides optimal performance, sparsity, and modularity. Notably, unlike fully connected MLP PINNs, we observed that second-order optimizers such as L-BFGS do not enhance accuracy in brain-inspired PINNs; they quickly converge to local minima without achieving higher precision.

Following the approach outlined in the seminal BIMT paper~\cite{liu2024seeing}, we divide the training into three phases with varying L1 penalization regimes: $\lambda$ starts at 0.001 with no bias penalization, increases to 0.01 to enhance locality at one-fourth of the total training, and finally, at three-fourths of the total training, switches to bias penalization while reducing $\lambda$ back to 0.001. We set $A$ to 2, and neural unit swaps occur every 200 epochs. Prior to the final error evaluation, we prune the PINN weights to eliminate weights and biases below $10^{-3}$ in absolute value.

For demonstration purposes, this work focuses on solving the one-dimensional Poisson equation with a harmonic source term:

\begin{equation}
\frac{d^2 x(t)}{dt^2} = \sin(t) + 4 \sin(2t) + 9 \sin(3t) + 16\sin(4t),
\label{problem}
\end{equation}

with boundary conditions $x(0) = 0$ and $x(2 \pi) = 0$ in the simulation domain $[0, 2 \pi]$. The Poisson equation is omnipresent in scientific computing: it is used for electromagnetics to solve electrostatic problems and in incompressible flow in computational fluid dynamics.

Regarding Eq.~\ref{problem}, This particular choice of harmonic source term allows us to evaluate different and higher spectral components. It has been observed that PINNs exhibit a spectral bias, converging rapidly to low-frequency parts of the solution while requiring more time to resolve high-frequency components accurately. By introducing various components with differing spectral characteristics, we can assess the performance of the brain-inspired architecture. We utilize 100 test collocation points to test and compare the results against the analytical solution. MSE and Euclidean error metrics are used throughout training to evaluate the performance of different PINN models.

The code utilized in this study is openly accessible on GitHub\footnote{\url{https://github.com/smarkidis/BrainInspiredPINN/}}.

\section{Results}
\label{results}
\begin{figure}[h!]
\centering
  \includegraphics[width= \textwidth]{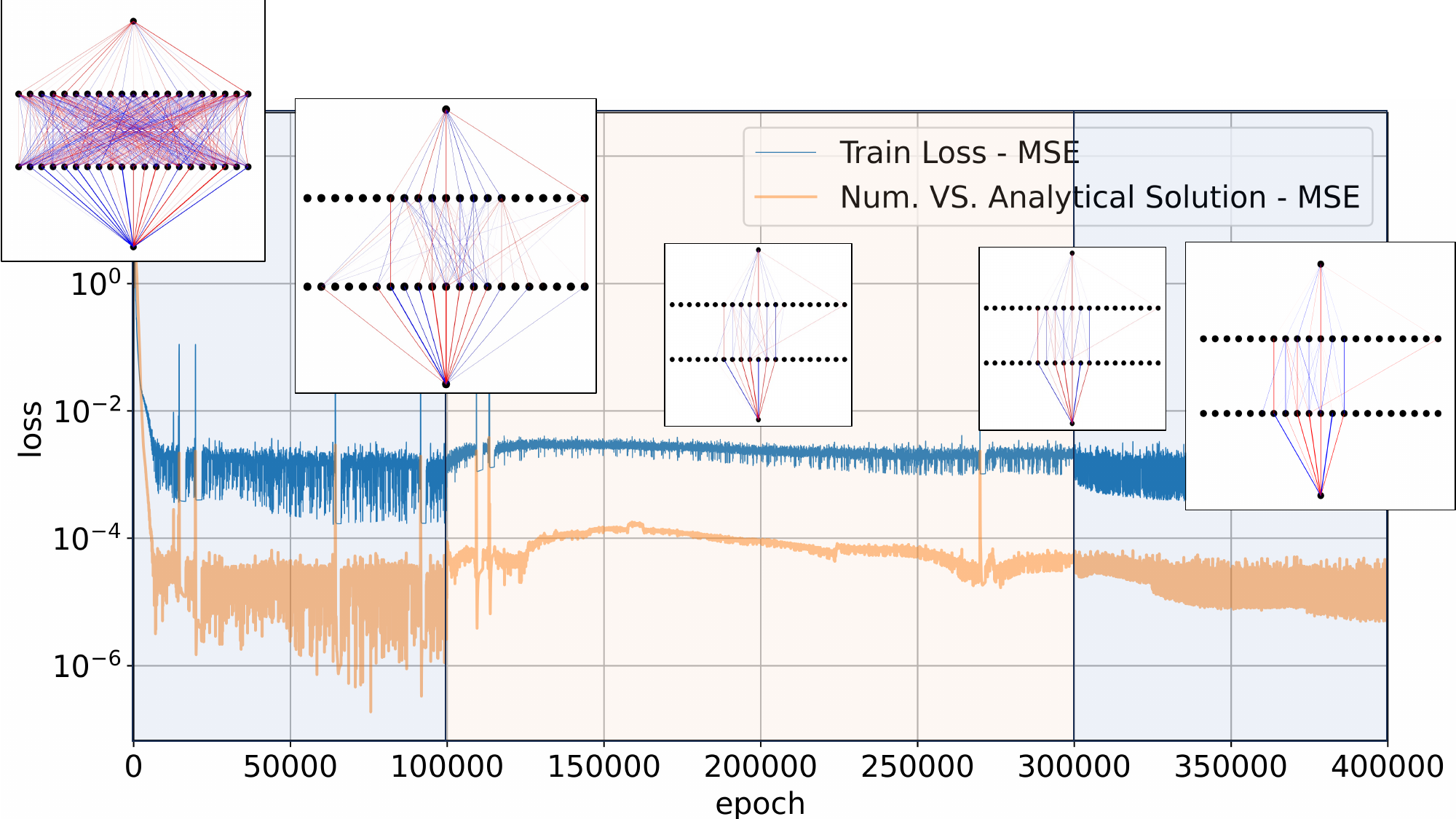}
  \caption{Evolution of the brain-inspired PINN network architecture during the training for the solution of $d^2 x(t) / dt^2 = \sin(t) + 4 \sin(2t) + 9 \sin(3t) + 16 \sin(4t)$ with $x(0) = 0$ and $x(2 \pi) = 0$.  The PINN architecture evolves from being fully connected to being highly sparse and modular. The red and blue lines represent connections associated with positive and negative weights. The training occurs in three phases where the strength of the L1 penalty (related to the importance of locality) changes.}
  \label{fig:loss}
\end{figure}

As the first step, we analyze the evolution of the brain-inspired PINN architecture during the training and assess its performance. In Fig.~\ref{fig:loss}, we show the results of the brain-inspired PINN network training applied to solve the 1D Poisson equation $d^2 x(t) / dt^2 = \sin(t) + 4 \sin(2t) + 9 \sin(3t) + 16 \sin(4t)$ with $x(0) = 0$ and $x(2 \pi) = 0$ for 400,000 epochs. The different inserts show the evolution of brain-inspired architecture. The red and blue edges connect neural units with positive and negative weights, respectively. We have evolved from a fully connected MLP PINN to a sparse and modular computer architecture, only utilizing a small part of the total capacity of the original network. In the background plot, the train loss and test error (calculated against the analytical solution) are represented in blue and orange colors.

By analyzing the losses, we can identify the three phases of the training as we change the value of $\lambda$ in the three phases. These three phases are represented as different background colors. Until the epoch of 100,000, the L1 penalty is relatively low, and the loss function decreases quickly; after that, we increase the importance of locality and note significant changes in the PINN architecture. At epoch 300,000, we decrease $\lambda$ and turn on bias penalization.

\begin{figure}
\centering
  \includegraphics[width= 0.6\columnwidth]{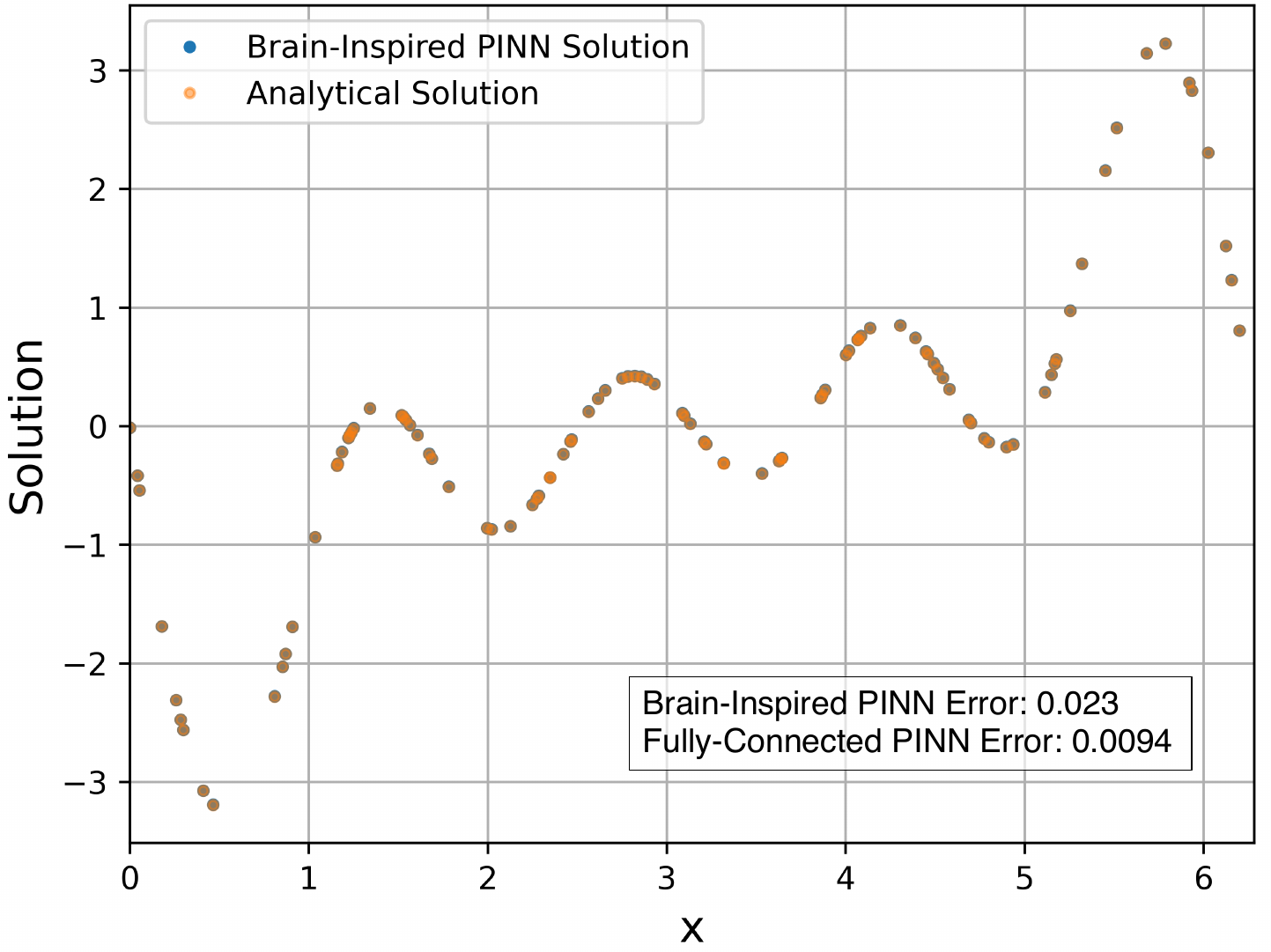}
  \caption{Analytical and brain-inspired solutions of the differential equation: $d^2 x(t) / dt^2 = \sin(t) + 4 s\in(2t) + 9 \sin(3t) + 16 \sin(4t)$ with $x(0) = 0$ and $x(2 \pi) = 0$. }
  \label{fig:solution}
\end{figure} 
The final PINN solution and its comparison with the analytical solution are presented in Fig.~\ref{fig:solution}. The brain-inspired PINN can capture all the frequencies present in the solutions at a reasonable accuracy. To estimate the loss of performance due to large sparsity and modularity, we also run a fully connected neural network with two hidden layers of 21 neural units and calculate the error in the Euclidean norm compared to the analytical solution. The final error for the brain-inspired and fully-connected PINNs are 0.023 and 0.0094, respectively. As pointed out, architecture with high sparsity comes with a performance loss: in this example, the brain-inspired PINN has approximately twice the error of the fully connected MLP PINN.

\subsection{Deriving Modular PINN Architectures}
\begin{figure}[h!]
\centering
  \includegraphics[width=\textwidth]{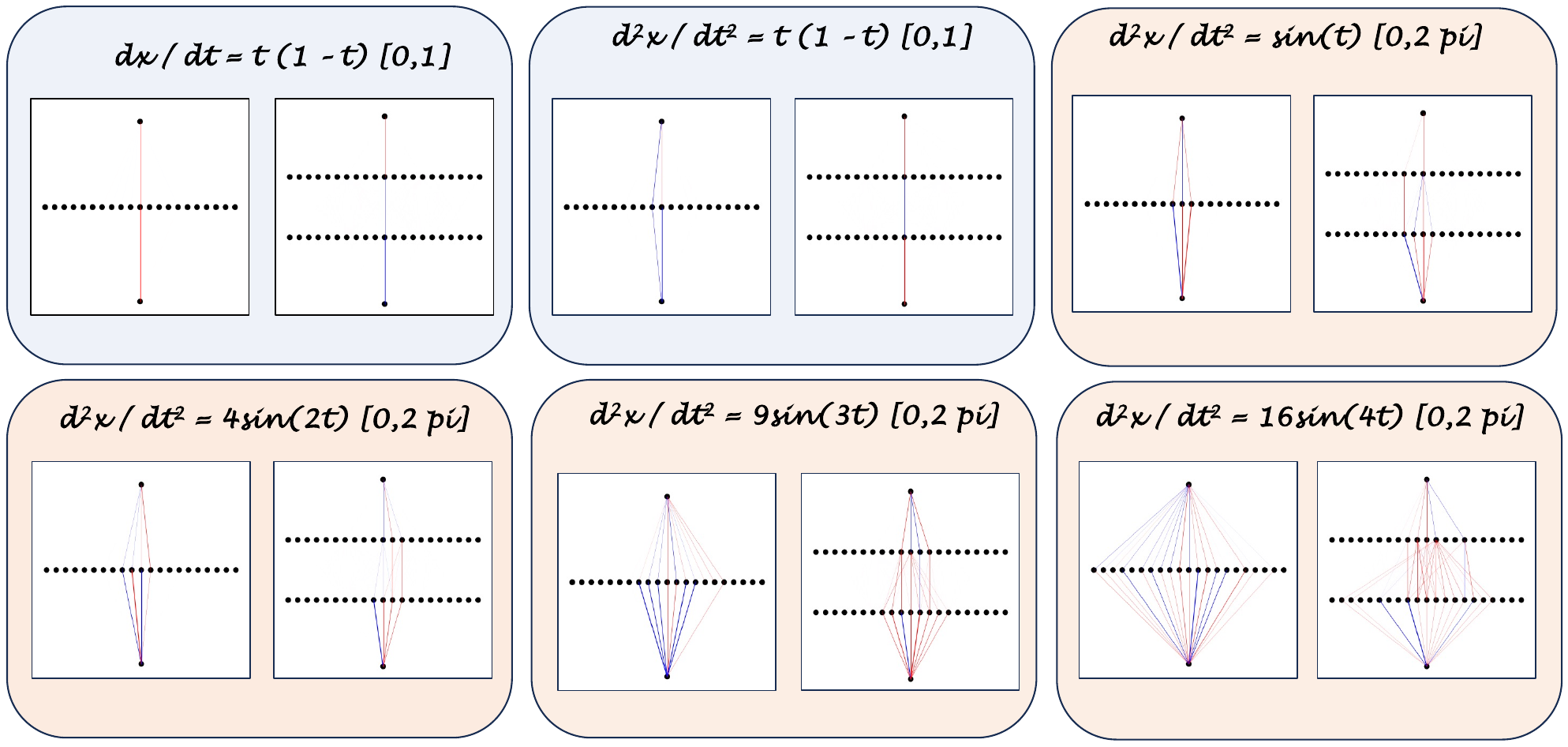}
  \caption{Bare-minimum architectures derived running brain-inspired PINN for solving basic differential equations. Interestingly, some differential equations, such as the logistic equation, only require a neural unit in the hidden layer. Another important point is that source terms with higher frequencies result in denser PINN architectures as a manifestation of the spectral bias phenomenon.}
  \label{fig:primitives}
\end{figure}

One of the advantages of brain-inspired neural networks is the possibility of deriving basic modules, the bare-minimum PINN architectures, that are small in scale and compact. These basic modules can be derived by solving simple problems, such as the logistic or Poisson equation in one dimension, with a simple archetypal source term, e.g., a sinusoidal source term with a single spectral component. Fig.~\ref{fig:primitives} shows different PINN bare-minimum architectures that can be derived by training brain-inspired PINN to solve simple archetypal problems with one and two hidden layers.

A few important points can be deduced by analyzing Fig.~\ref{fig:primitives}. For instance, it is striking that the solution of the logistic equation (top left panel of Fig.~\ref{fig:primitives}) requires only one neural unit in the hidden layer. Another important point comes up when analyzing the final brain-inspired PINN architecture obtained by training the neural network for source terms with higher frequency terms: to solve a low-frequency signal $ \sin(t)$ requires only three neural units in the hidden layer. As we increase the frequency of the source term, we note that the number of neural units in the hidden layer increases. For instance, when solving the one-dimensional Poisson equation with  $ 16\sin(4t)$, the brain-inspired converges to a fully connected MLP. In general, we observe that by increasing the frequency of the source term, more neural units are needed in the hidden layer to converge to a solution. This clearly manifests the spectral bias in the number of neural network connections needed to express higher-frequency components. Using a brain-inspired approach, we show that spectral bias is not only in the rate of convergence but also in the architecture of the PINN: higher frequency signal requires a larger number of neural units and layers to be accurately resolved. 

As an additional note, we remark that bare-minimum architectures depend on the activation function that provides the basic basis function to express the solution. Using other activation functions leads to slightly different architecture modules for the problem presented here.

An essential aspect of the experiments shown in Fig.~\ref{fig:primitives} is that we can identify PINN bare-minimum architectures that we can use in larger architectures. These basic building blocks can be used similarly to convolutional~\cite{lecun1995convolutional} and attention~\cite{vaswani2017attention} modules in established deep neural networks. Here, we can utilize a modular architecture based on the module derived from solving the Poisson equation in 1D for the $\sin(t)$ source term. For instance, we can use three modules as depicted on the right side of Fig.~\ref{fig:teaser}: the overall neural network consists of three building blocks with the same collocation point as input and whose output is combined by summing up the output of each building block. The modular PINN combines the modules into a larger NN using \texttt{PyTorch} superclasses and inheritance and is trained as traditional PINNs with an Adam optimizer. This is a simple example of modular PINN architecture, and larger modular architecture can be obtained by using a larger number of modules or using more advanced modules, e.g., modules obtained for two or more hidden layers or different source terms.

\begin{figure}[h!]
\centering
  \includegraphics[width= 0.75 \textwidth]{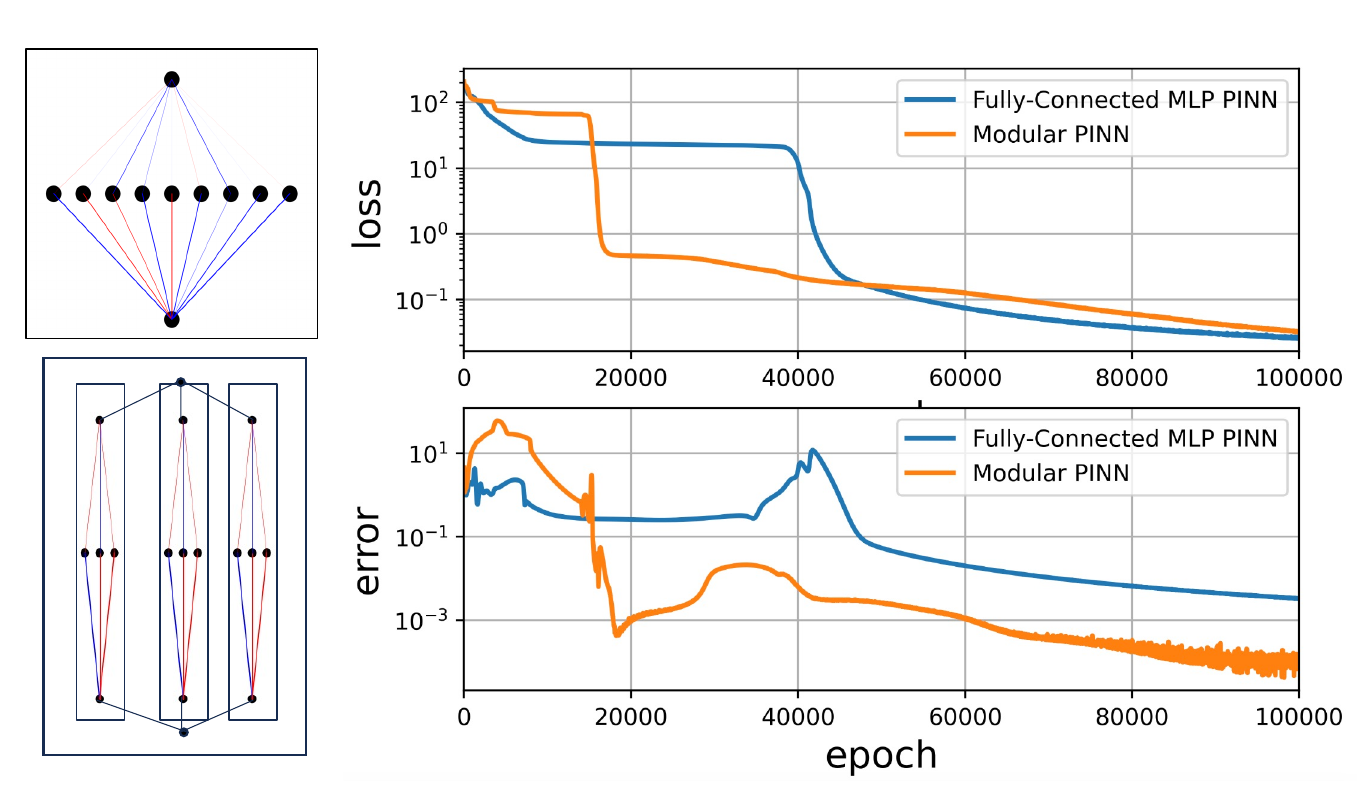}
  \caption{Performance comparison between a fully connected MLP network with a hidden layer with nine neural units and a modular PINN using a network primitive obtained by solving $d^2 x(t) / dt^2 = \sin(t)$. The modular PINN has three basic modules with three neural units. While the fully connected and modular PINNs have similar test loss values, the MSE against the analytical solution is two orders lower for the modular PINN.}
  \label{fig:modular}
\end{figure}

To understand the potential benefit of the modular compact PINN architecture, we solve the original problem expressed by Eq.~\ref{problem} with a fully connected MLP PINN with one hidden layer and nine neural units; we compare the loss for the training data set and the error of the solution against the analytical solutions. We present the results in Fig.~\ref{fig:modular}.

When analyzing Fig.~\ref{fig:modular}, we note that the training loss values are similar, showing a comparable performance. However, the results of the two network architectures, regarding the test error and comparison of the analytical solution (bottom panel of Fig.~\ref{fig:modular}), show a considerably higher performance of the modular PINN. We can see that the modular PINN has a test error (calculated as MSE) that is approximately two orders lower than the fully-connected MLP PINN. Compared to the analytical solution, the test dataset's final Euclidean error is 0.083 for the modular PINN vs. 0.57 for the fully-connected MLP PINN. We also note that a simple modular neural network performs better than the PINN networks (including the brain-inspired NN), whose results are previously presented in Fig.~\ref{fig:solution}. The modular PINN, built on the top of a module obtained with BIMT, exhibits improved generalization properties in this use case and compact formulation.

\section{Related Work}
 \label{related}
This work is at the intersection of PINN, brain-inspired neural networks, and modular structures for neural networks. PINNS have been extensively studied since their introduction by Raissi et al. in Ref.~\cite{raissi2019physics}. Subsequent research has focused on aspects such as convergence, stability, and numerical properties. PINNs have demonstrated versatility across various applications, including computational fluid dynamics~\cite{jin2021nsfnets}, solid mechanics \cite{haghighat2021physics}, molecular dynamics \cite{hassanaly2023pinn}, and battery life cycle modeling~\cite{hassanaly2023pinn}. However, previous PINN studies have primarily focused on fully connected architectures without incorporating sparsity or modularity.

Brain-inspired machine learning is gaining prominence as researchers seek inspiration from neural computations. Modular architectures have been a key focus in this domain. Convolutional Neural Networks (CNNs)~\cite{lecun1995convolutional} exemplify modular architecture, preserving symmetries and invariances under various transformations. Similarly, transformer neural networks~\cite{vaswani2017attention} utilize attention mechanisms as fundamental building blocks. Additionally, graph neural networks \cite{bronstein2017geometric} demonstrate structured architectures for processing graph data. These modular approaches enhance interpretability and scalability in neural network design.

Unlike previous PINN studies, this work introduces a novel approach where the PINN architecture evolves into a sparse and modular structure. By integrating brain-inspired techniques and modular design principles, this study explores new avenues for enhancing PINN architectures, potentially improving efficiency and interpretability in solving partial differential equations.

\section{Discussion and Conclusion}
 \label{conclusion}
This study investigated a fundamental aspect of PINN architectures, widely employed for solving differential equations. Conventionally, PINN methodologies have predominantly relied on fully connected MLP architectures. However, traditional numerical solvers for differential equations exhibit sparsity and a modular structure, wherein computations at a given point depend only on neighboring points, termed the stencil. This work merges established PINN techniques with a brain-inspired neural network approach to address these architectural limitations and enhance the solution of differential equations. Specifically, we leverage brain-inspired neural networks to achieve two primary objectives: first, to derive basic building blocks for constructing larger neural networks, and second, to obtain solutions for differential equations.

From a computational perspective, brain-inspired neural architectures offer significant advantages in terms of sparsity, leading to reduced computation and memory requirements. The examples presented in this work demonstrate remarkable sparsity, with simple equations requiring minimal neural units to achieve satisfactory accuracy. However, the degree of sparsity is influenced by various factors, including the characteristics of the differential equations, particularly those involving high-frequency components. While the achieved sparsity impacts memory storage needs, current support for sparse computations in mainstream deep-learning frameworks, such as PyTorch and TensorFlow, remains limited, necessitating advancements in this area to exploit computational benefits fully. Usage of frameworks and libraries for NN computation~\cite{mishra2021accelerating,ivanov2023sten}, directly in sparse formats, such as CSR/CSC/COO~\cite{langr2015evaluation}, is key to achieving both computational advantages, lowering the complexity costs of matrix multiply and memory requirements.

We observed that by increasing the frequency of the source term, more neural units are needed in the hidden layer to converge to a solution. This clearly manifests the spectral bias in the number of neural network connections needed to express higher-frequency components. Using a brain-inspired approach, we show that spectral bias is visible not only in the rate of convergence but also in the architecture of the PINN: higher frequency signal requires a larger number of neural units and layers to be accurately resolved. Understanding this phenomenon clarifies the challenges inherent in training PINNs for problems with high-frequency components and indicates the importance of developing architectures capable of accommodating such spectral biases.

This study proposed a novel contribution to PINNs architectures by introducing an approach to constructing modular architectures. By leveraging brain-inspired neural network techniques, we derived PINN bare-minimum architectures through BIMT on basic archetype problems. These modular building blocks (the bare-minimum architectures) exhibit potential for application in larger architectures reminiscent of the modular structures found in convolutional and attention-based modules within established deep NN. Modular PINNs offer promising possibilities for increasing the accuracy and efficiency of solving PDE problems, bridging the gap between traditional numerical methods and machine learning approaches~\cite{markidis2021old}. Moreover, adopting modular architectures facilitates a transition from complex, fully connected Multilayer Perceptron (MLP) designs to simpler, more compact PINN architectures, thus reducing computational overhead and memory requirements. This study clarifies the architectural evolution of PINNs. It prepares for future research into developing and optimizing modular neural network frameworks for a wide range of scientific and engineering applications.

However, it is important to acknowledge that this study represents only an initial exploration of modular PINN architectures. Our investigation focused primarily on simple one-dimensional problems and utilized only a single building block derived from solving the Poisson equation with a sinusoidal source for the modular PINN. While our findings are promising, further research is needed to refine the composition of these building blocks and optimize the accuracy of compact and modular PINN architectures. Future studies could explore the incorporation of additional architectural primitives derived from a broader range of differential equations and problem domains. Additionally, research on enhancing the modularity and versatility of these architectures by incorporating multi-layer modules and exploring alternative activation functions could further improve their performance.




\end{document}